\documentclass{ws-procs975x65}

\begin{document}

\title{ PNS DYNAMO: THEORY AND OBSERVATIONS}

\author{ALFIO BONANNO}
\address{
INAF, Osservatorio Astrofisico di Catania,
        Via S.Sofia 78, 95123, Catania, Italy\\
INFN, Sezione di Catania,
        Via S.Sofia 72, 95123, Catania, Italy}

\author{VADIM URPIN}
\address{
   A.F. Ioffe Institute of Physics and Technology \\
Isaac Newton Institute of Chile, Branch in St. Petersburg,
194021 St. Petersburg, Russia}

\begin{abstract}
We briefly review the turbulent mean-field dynamo action in protoneutron
stars that are subject to convective and neutron finger
instabilities during the early evolutionary phase. 
By solving the mean-field induction equation with the simplest model
of $\alpha$-quenching we estimate the strength of the generated
magnetic field. 
If the initial period of the protoneutron star is short,
then the generated large-scale field is very strong ($> 3 \times
10^{13}$G) and exceeds the small-scale field at the neutron star
surface, while if the rotation is moderate, then the pulsars are formed with
more or less standard dipole fields ($< 3 \times 10^{13}$G) but
with surface small-scale magnetic fields stronger than the dipole
field. If rotation is very slow, then the mean-field dynamo does not
          operate, and the neutron star has no global field. 
\end{abstract}


\bodymatter

\section{Introduction}\label{aba:sec1}
A protoneutron star (PNS) is a very hot ($T\sim 10^{11} \; K$), rapidly rotating, lepton rich object
that has been formed from the collapse of a massive stellar progenitor.
It is believed that lepton and negative entropy gradients generates 
hydrodynamical instabilities which can play a significant dynamical role in the 
early stage of the PNS evolution.
(Grossman et al.  1993, Bruenn \& Dineva 1996, Miralles et al. 2000). 
While convective instability is 
presumably connected to the entropy gradient, the so-called 
neutron-finger instability is instead generated by a negative lepton gradient. 
The latter is due to dissipative processes which 
are rather fast in PNSs and it grows on a timescale $\sim 
30-100$ ms, that is one or two orders of magnitude longer than the 
growth time of convection (Miralles et al. 2000). Turbulent motions 
caused by hydrodynamic instabilities in combination with rotation 
make turbulent dynamo one of the most plausible mechanism of the pulsar magnetism.

The character of turbulent dynamo depends on the Rossby number, $Ro = 
P/ \tau$, where $P$ is the PNS spin period and $\tau$ the turnover 
time of turbulence. If $Ro \gg 1$, the effect of rotation on turbulent 
motions is weak and the mean-field dynamo is inefficient. The small-scale 
dynamo can be operative, however, even at very large Rossby numbers. If 
$Ro \leq 1$ and the turbulence is strongly influenced by the rotation, then the 
PNS can be subject to a large-scale  mean-field dynamo action. Note that the 
small-scale dynamo can still operate in this case. The Rossby number is 
typically large in the convective region, $Ro \sim 10-100$, and the 
mean-field dynamo is not likely to work in this region. On the contrary, 
except very slowly rotating PNSs, the Rossby number is of the order of unity,
$Ro \sim 1$, in the neutron-finger unstable region 
(Bonanno et al. 2003,2005) where turbulent motions are 
slower, and the turbulence is strongly modified by rotation. This favors 
the efficiency of mean-field dynamos in the neutron-finger unstable 
region. This dynamo mechanism is then very different from the one proposed
by Thompson \& Duncan (1993) who argued that only small-scale dynamos can operate in most PNSs. 

\def\figsubcap#1{\par\noindent\centering\footnotesize(#1)}
\begin{figure}[b]%
\begin{center}
%  \parbox{2.1in}{\epsfig{figure=1.ps,width=1.8in}\figsubcap{a}}
  \parbox{2.1in}{\epsfig{figure=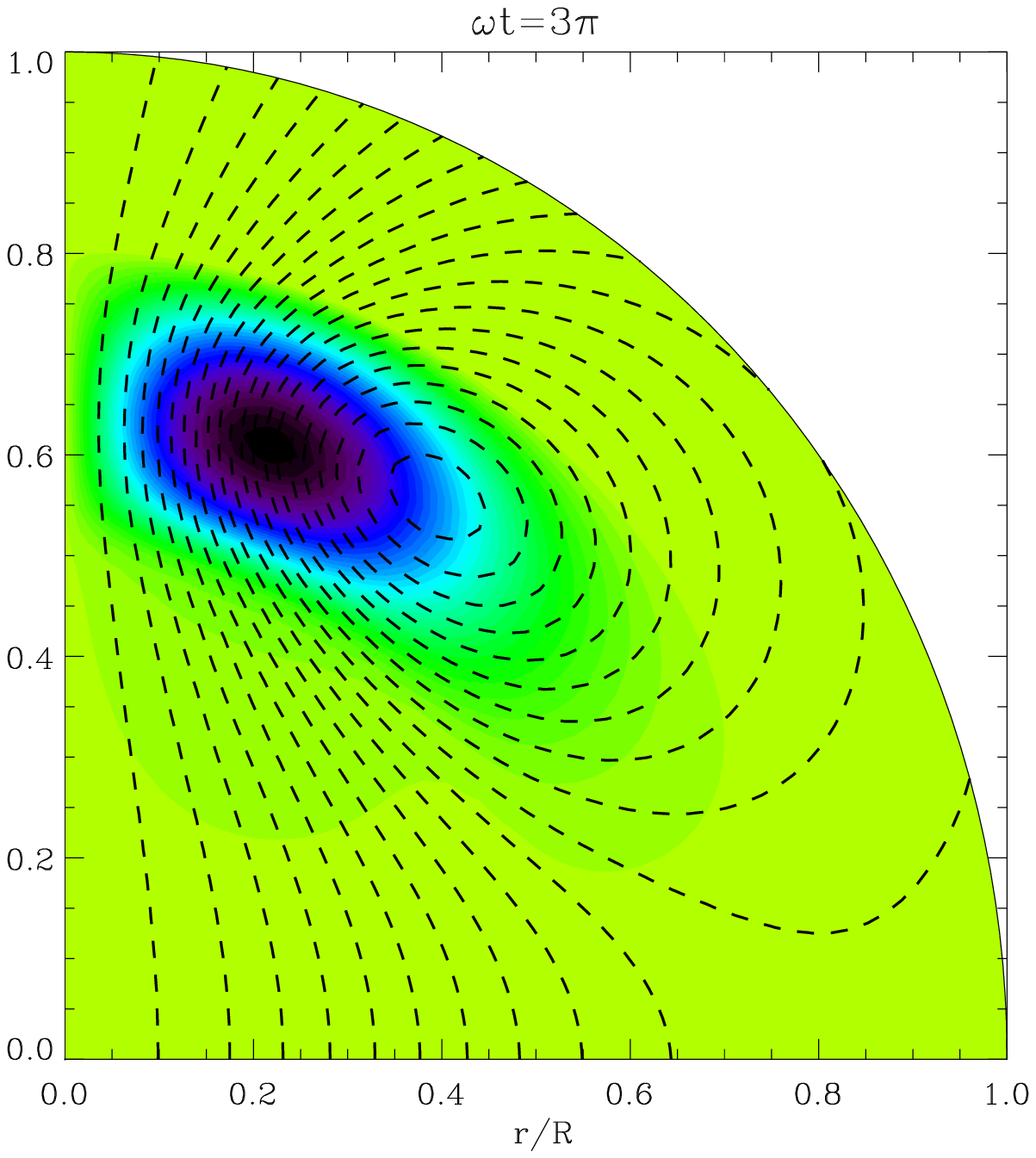,width=1.8in}}
  \hspace*{4pt}
%  \parbox{2.1in}{\epsfig{figure=2.ps,width=1.8in}\figsubcap{b}}
  \parbox{2.1in}{\epsfig{figure=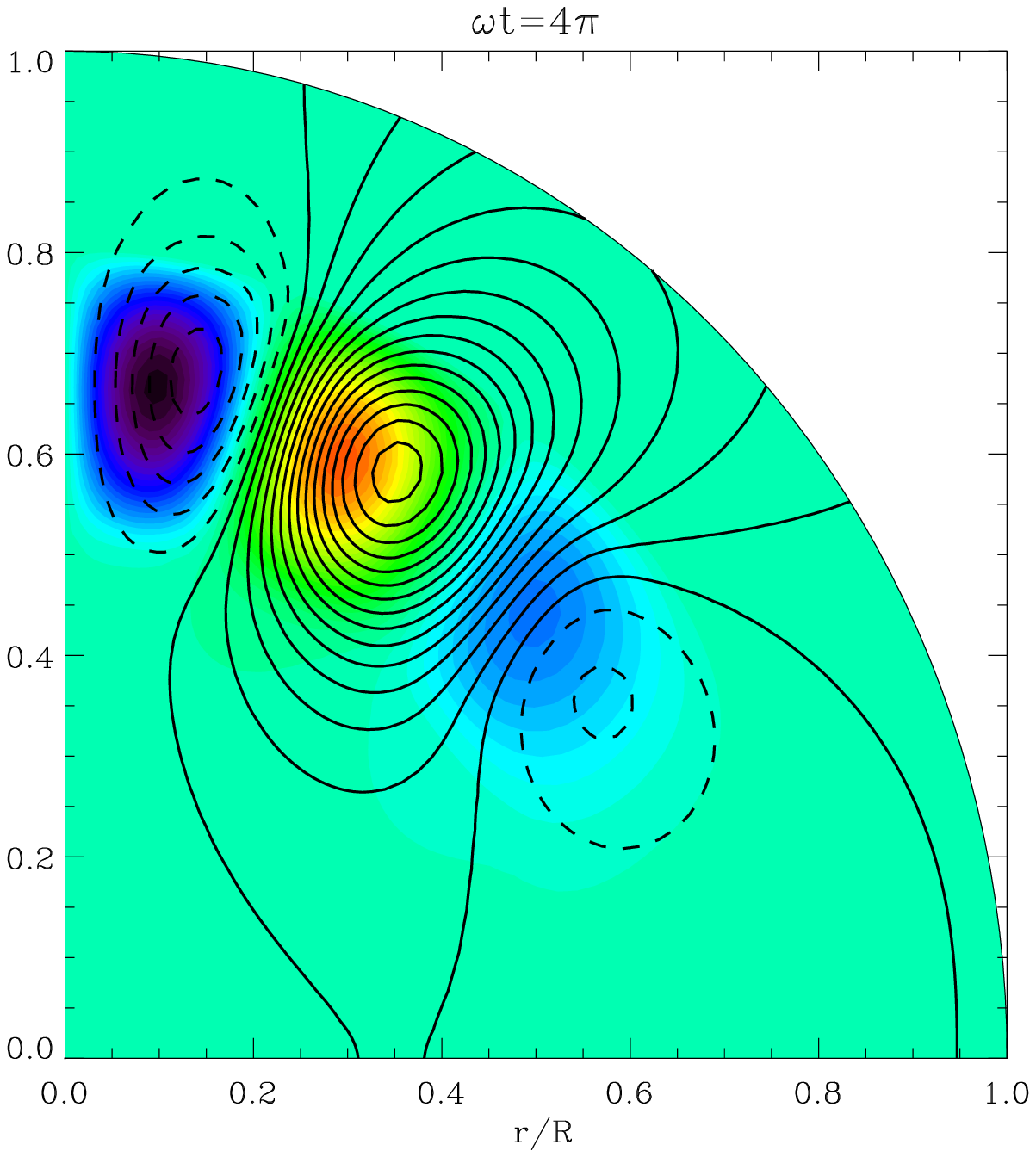,width=1.8in}}
  \caption{On the left panel is depicted a typical field configuration for $\alpha^2$-dynamo
while in the right panel is depictied a field configuration in the presence of differential rotation.
Color levels are for the toroidal field, while dashed lines represents poloidal field lines.  }%
  \label{fig1.2}
\end{center}
\end{figure}
\section{The model }
The problem for dynamo modelling in PNSs has been described by Bonanno et al. 
(2003, 2005). We model the PNS as a sphere of radius $R$ with two different 
turbulent zones separated at $R_{c}<R$. The inner 
part ($r<R_{c}$) corresponds to the convective region, while the outer 
one ($R_{c}<r<R$) to the neutron-finger unstable region. The mean-field 
equation reads 
\begin{equation}
\frac{\partial \vec{B}}{\partial t} = \nabla \times (\vec{v} 
	\times \vec{B} + \alpha \vec{B})
	- \nabla \times (\eta \nabla \times \vec{B}) \ ,
\end{equation}
where $\eta$ and $\alpha$ are the turbulent magnetic diffusivity and 
$\alpha$-parameter of the dynamo theory, respectively (For details see Bonanno et al. 2006). 
We assume that the rotation is the only large-scale motion and $\vec{v} = \vec{\Omega}(\vec{r}) \times 
\vec{r}$. We can basically distinguish two different dynamo regime: in the $\alpha-\Omega$ regime the dynamo action
is mainly due to the differential rotation, while in the $\alpha^2$ regime the dynamo action is mainly 
due to the turbulent helicity. We found that most of the PNS do generate a large scale mean field dynamo provided
the initial period is shorter than a critical period $P_c$ which is of the order of $0.1-1 \rm s$ (See Bonanno et al. 2003). 
The typical field configuration in presence of differential rotation is depicted in Fig.1 where we have 
supposed that the boundary of the NF unstable region is located at 0.6 stellar radii. 
As it is possible to note the region of the maximum strength of the generated field is inside the region of instability. 
\section{Results}
The critical period that determines the onset of the mean-field dynamo action is 
rather long, and dynamos should be effective in most PNSs. The unstable 
stage lasts $\sim 40$ s, and this is sufficient for the dynamo to 
reach saturation. We can estimate a saturation field assuming 
the simplest $\alpha$-quenching with non-linear $\alpha$ given by
%\begin{equation}
$\alpha (\tilde{B}) = \alpha_{nf} (1 + \tilde{B}^{2}/B_{eq}^{2})^{-1}$,
%\end{equation}
where $\tilde{B}$ is the characteristic value of the generated field,
and $B_{eq}$ is the equipartition small-scale magnetic field. The generated 
field reduces the $\alpha$-parameter and slows down the generation. The 
saturation is reached when non-linear $\alpha$ becomes equal to the critical 
value $\alpha_{0}$ corresponding to the marginal dynamo stability. Then, the 
saturation field is
%\begin{equation}
$B_{s} \approx B_{eq} \sqrt{P_{c}/P-1}$ .
%\end{equation}
A subsequent activity of the PNS as a radiopulsar is determined by the 
poloidal component of the saturation field, $B_{ps}$ and by the ratio $\xi = |B_t/B_p|$. 
Using the
estimate $B_{s} \sim B_{ps} (1 + \xi)$, for $B_{ps}$ we obtain
%\begin{equation}  
$B_{ps} \approx B_{eq} (1 + \xi)^{-1} \; \sqrt{P_{c}/P-1}$,
%\end{equation}
The equipartition field, $B_{eq} \approx 4 \pi \rho v_{T}^{2}$, varies 
during the unstable stage, rising rapidly soon after collapse, reaching 
a quasi-steady regime, and then going down when the temperature and lepton 
gradients are smoothed. We can estimate $B_{eq}$ at a peak of instabilities
as $\sim 10^{16}$ G in the convective zone, and $\sim (1-3) \times 10^{14}$ 
G in the neutron-finger unstable zone (Urpin \& Gil 2004). However, the 
temperature and lepton gradients are progressively reduced as the PNS cools 
down, hence, $v_{T}$ and $B_{eq}$ decrease whereas the turnover time 
$\tau$ increases. The meand field dynamo description can be applied as the quasi-steady condition 
$\tau_{cool} \gg \tau$ is fulfilled, $\tau_{cool}$ being the cooling timescale. 
The final equipartition field is the same for the both the unstable zones.
We thus have $B_{eq} \sim (1-3) \times 10^{13}$ G for the largest 
turbulent scale ($\ell_{T} =L \sim 1-3$ km) if $\tau_{cool} \sim$ few 
seconds. 

We can distinguish few types of neutron stars which 
possess different magnetic characteristics: (i) 
{\it Strongly magnetized neutron stars.} If the initial period satisfies 
the condition $P < P_{m} \equiv P_{c} [1 + (1 + \xi)^{2}]^{-1}$ then the 
dynamo action leads to the formation of a strongly magnetized PNS with $B_{ps} 
> B_{eq} \sim 3 \times 10^{13}$G. Since $Ro \leq 1$ for fast rotators, 
we can expect that the rotation of such stars is almost rigid. For a rigid 
rotation, the $\alpha^{2}$-dynamo is operative, hence, $\xi \sim 1-2$ 
in the neutron-finger unstable region. Therefore, strongly magnetized stars 
can be formed if the initial period is shorter than $\sim 0.1 P_{c} $. 
%\varepsilon P_{0}$ being $\varepsilon =l_{\rm L}^2/L^2 \sim 1$. Then for the surface field of such PNSs we have
Then for the surface field of such PNSs we have
%\begin{equation}
$B_{ps} \sim 0.3 B_{eq} \sqrt{P_{c}/P -1 } \sim 10^{13}\sqrt{P_{c}/P -1} \;\;\; {\rm G}$.
%\end{equation}   
The shortest possible period is likely $\sim 1$ ms, hence, the strongest 
mean magnetic field that can be generated by the dynamo action is $\sim 3 
\times 10^{14}$ G. 
%This value is only a bit higher than the maximal field 
%inferred from the spin-down data in ragiopulsars, $\sim 9.4 \times 10^{13}$ G 
%(McLaughlin et al. 2003). Likely, the field of all five high-magnetic-field 
%radiopulsars known at the moment (PSR J1847-0130 ($9.4 \times 10^{13}$ G), PSR 
%J1718-3718 ($7.4 \times 10^{13}$ G), PSR J1814-1744 ($5.5 \times 10^{13}$ G), 
%PSR J1119-6127 ($4.4 \times 10^{13}$ G), and PSR B0154+61 ($2.1 \times 
%10^{13}$ G)) has been generated in this regime, and these neutron stars 
%were very fast rotators with $P$ of the order of few ms at their birth. Note that some 
%PNSs, which had the initial period $P < 0.1 \varepsilon P_{0}$ and were 
%strongly magnetized just after the unstable stage, now may possess a 
%dipole field $< 3 \times 10^{13}$ G because a fraction of the electric 
%current decays during the early evolution when the conductivity is relatively 
%low (Urpin \& Gil 2004). 
Since the small-scale magnetic field is weaker than 
the large-scale field at the surface of such stars, one can expect 
that radiopulsations from them may have a more regular structure than those 
from  low-field pulsars.
%The maximum field generated in rapid rotating PNSs is comparable to 
%estimates of the surface field in some SGRs and AXPs which are believed to 
%be the candidates in magnetars. However, estimates of $B$ inferred from 
%spin-down are not reliable for AXPs and SGRs because $\dot{P}$ 
%can vary significantly on a short timescale that seems to be unrealistic 
%for the magneto-dipole braking (Kaspi \& McLaughlin 2004).   
(ii) {\it Neutron stars with moderate magnetic fields.} If $P_{c} > P > 
P_{m}$ then the generated mean poloidal field (for example, dipolar) is 
weaker than the small-scale field, $B_{ps} < B_{eq} \sim 3 \times 
10^{13}$G. The Rossby number is $\sim 1$ or slightly larger for such PNSs, 
and their rotation can be differential. Departures from a rigid rotation 
are weak if $P$ is close to $P_{m}$ but can be noticeable for $P$ close to
$P_{c}$. As a result, the parameter $\xi$ can vary within a wider range 
for this group of PNSs, $\xi \sim 5-60$, and the toroidal field should be 
essentially stronger than the poloidal one in the neutron-finger unstable 
region. The toroidal field is stronger than the small-scale field if 
$P_{c}/2 > P > P_{m}$, and weaker if $P_{c} > P > P_{c}/2$. Note that both 
differential rotation and strong toroidal field can influence the thermal 
evolution of such neutron stars. Heating caused by the dissipation of the
differential rotation is important during the early evolutionary stage 
because viscosity operates on a relatively short timescale $\sim 10^{2}-
10^{3}$ yrs. On the contrary, ohmic dissipation of the toroidal field is 
a slow process and can maintain the surface temperature $\sim (1-5) \times 
10^{5}$K during $\sim 10^{8}$ yrs (Miralles et al. 1998). 
%The magnetic field of this type of neutron stars should be very irregular 
%with a number of sunspot-like magnetic structures on their surface. This 
%concerns in particular  relatively slowly rotating PNSs with $P$ slightly 
%shorter than $P_{c}$ where only a weak dipole magnetic field can be 
%generated. As it was argued by urpin \& Gil (2004), magnetic spots with 
%the lengthscale $\gtrsim 1$ km can survive during the whole lifetime of 
%radiopulsars. One can expect, therefore, that radiopulsations from this type
%of pulsars may have a complex structure. For example, small-scale magnetic 
%structures can be responsible for the phenomenon of drifting subpulses 
%observed in many pulsars (Gill \& Sendyk 2003). Note also that features in 
%the X-ray spectra of this group of pulsars may indicate the strength of the 
%magnetic field which differs essentially from the field strength inferred 
%from the spin-down data. This can happen because spectral features provide 
%information about the strength of small-scale fields at the surface rather 
%than about the mean field which is associated to the magneto-dipole braking.  
(iii) {\it Neutron stars with no large-scale field.} If the initial period is
longer than $P_{c}$ then  a large scale mean-field dynamo does not 
operate in the PNS but the small-scale dynamo can still be efficient. We
expect that such neutron stars have only small-scale fields with the 
strength $B_{eq} \sim 3 \times 10^{13}$ G and no dipole field. Likely, 
such slow rotation is rather difficult to achieve if the angular momentum 
is conserved during the collapse, and the number of such exotic PNSs is small. 

Likely, the most remarkable property of these neutron stars is a discrepancy
between the magnetic field that can be inferred from  spin-down
measurements and the field strength obtained from  spectral observations.
Features in X-ray spectra may indicate the presence of rather a strong
magnetic field $\sim 3 \times 10^{13}$ G (or a bit weaker because of ohmic
decay during the early evolution) associated to sunspot-like structures at
the surface of these objects. The field inferred from  spin-down data
should be essentially lower.

In particular the recent discovery of young slowly spinning radio pulsar 
1E 1207.4-5209 (Gotthelf \& Halpern  2007) with a very weak magnetic field 
seems to support our theory.  

{}


\begin{thebibliography}{99}

\bibitem{}
Bonanno A., Rezzolla L., Urpin V. 2003. A\&A, 410, L33

\bibitem{}
Bonanno A., Urpin V, Belvedere G. 2005. A\&A, 440, 149 

\bibitem{}
Bonanno A., Urpin V, Belvedere G. 2006. A\&A, 451, 1049 

\bibitem{}
Bruenn S., Dineva T. 1996, ApJ, 458, L71

\bibitem{}
Gotthelf E.V., Halpern J.P., 2007, ApJ, 664L, 35

\bibitem{}
Miralles J., Pons J., Urpin V. 2000, ApJ, 543, 1001 

\bibitem{}
Thompson C., Duncan R. 1993, ApJ, 408, 194

\bibitem{}
Urpin V., Gil J. 2004. A\&A, 415, 305

\end{thebibliography}
\end{document}